\documentclass[prb,preprint]{revtex4-1} 

\usepackage{amsmath}  
\usepackage{amsfonts} 
\usepackage{graphicx} 
\usepackage{float}
\usepackage[normalem]{ulem}
\usepackage{cancel}
\usepackage[usenames,dvipsnames]{color} 
\usepackage{amsmath} 
\usepackage{amssymb, bm}
\usepackage{soul} 
\usepackage{ifthen} 
\usepackage{upgreek}
\usepackage[ampersand]{easylist}
\usepackage{todonotes}
\usepackage{transparent}
\usepackage{xifthen}
\usepackage{xcolor}
\usepackage{verbatim}
\usepackage[define-L-C-R]{nicematrix}
\begin{document}


\title{Novel, first-principles approach to deriving Electromagnetic field transformations under oblique Lorentz-Boosts and arbitrary Spatial Rotations}

\author{Salil K. Bedkihal}
\affiliation{\mbox{Independent}}
	
\author{Rajeev K. Pathak}
\affiliation{Department of Physics, S.P. Pune University, Pune 411007, Maharashtra State, India and
Tulane University, New Orleans, Louisiana 70118, U.S.A. }

\date{ \today}




\date{\today}
\begin{abstract} 
The standard classic special-relativistic transformation of the electromagnetic (EM) field under proper Lorentz transformations is revisited. As to the pure Lorentz-boosts, popular treatments on EM transformation contemplate ideal geometries generating special static charge
and steady current distributions and in conjunction, invoke parallel and perpendicular (to the boost-velocity) components of the fields so engendered; the outcomes subsequently being
suitably “generalized”. We demonstrate {\it ab initio}, the EM-field transformation under arbitrary oblique uniform relative boost velocities from a broader, rigorous yet remarkably lucid perspective, exploiting the anti-symmetry of the EM-field tensor and symmetry of the general Lorentz-boost transformation— preemptively eliminating the labor of performing complicated matrix multiplications. Gratifyingly, this instructive exercise manifestly encompasses time varying fields and yields the transformation relations in a coordinate-free form. Further, the EM field transformation under proper, instantaneous 3-D rotations is demonstrated to transform as a
passive rotation of the coordinate axes, where the electric and magnetic fields transform
independently as purely electric and magnetic fields, respectively. The present didactical exercise, amenable to be incorporated in graduate courses on relativistic electrodynamics, should serve as an excellent illustration on how powerful symmetry arguments simplify the otherwise tedious calculations while endowing complete generality to the connections obtained.
\end{abstract}

\maketitle 
\section{prologue} 
The accompanying paper re-visits a celebrated, well-known text-book-result in classical
electrodynamics, on how does the electromagnetic (EM) field transform between two ‘Lorentz’
frames of reference under oblique Lorentz-boosts and proper rotations. As it stands, practically
every standard text book on electrodynamics discusses the former only under particular
special scenarios. The following legitimate (and obvious) questions may then be asked, the
answers to which form the very propriety in re-visiting the familiar, established result.

 {\bf Is the result derived herein (Transformation of the Electromagnetic field under
Lorentz boosts and Rotations) well-known?}

 ---Yes. The result discussed is a standard, classic text-book result that practically all the
standard texts attempt to “derive”, at least as to the boost-part.

 {\bf Why re-visit the classic result, then? What is the element of “novelty” in this work?} 
 
 \textendash\textendash The crucial result ought to be derived on firm, general grounds from first principles,
hence the need to re-visit. The novelty is essentially in the methodology, in the generality of the
rigorous proof, without taking recourse to particular scenarios: The standard texts on
electrodynamics DO NOT derive the result from first principles. They obtain the EM-field
transformation employing a ‘recipe’, by contemplating certain ideal, special static charge and
steady current distributions with preferred symmetries and chosen geometries. Next, ‘parallel’
or, longitudinal and ‘perpendicular’ or transverse —to the boost-velocity, conveniently chosen
along a chosen Cartesian coordinate axis— components of the static fields so generated are
invoked. The EM-field connections are then obtained for only these components and the
resulting special EM-field transformation equations are suitably ‘generalized’.
 This procedure manifestly has lacunas: (i) First, it is not a first-principles (‘ab initio’)
derivation. (ii) Second, with actual dynamic (time-varying) sources, i.e. charge and current
distributions, the procedure becomes irresolute and hence inapplicable. The result is assumed to
hold ‘on faith’. (iii) Third, if the source-geometries are completely unsymmetrical, and
moreover, (iv) if the counterpart boosted inertial frame moves with a general oblique relative
boost velocity, the ‘parallel-perpendicular’ strategy fails, thus becomes untenable.
‘Generalizing’ from particular instances, is awkward.
 Toward an affirmative end, however, a most general derivation is indeed possible as
presented herein. This is accomplished, essentially, by appealing to the symmetry properties of
the general Lorentz-boost transformation (with an oblique relative velocity) matrix and the antisymmetry of the electromagnetic field tensor, leading to an immaculate, ab initio, ingenious
general derivation, to which this article devotes itself in its first part, while transformation under
an arbitrary, pure, 3-D rotation forms the theme of the latter segment.
 Thus, the accompanying submission presents an entirely novel and general approach to the
well-known —but-never-cleanly-derived— classic problem : the merit of the article should be
gauged in the spirit of its providing a novel and clear-cut general proof for a well 
established crucial result that (classically) unifies the phenomena of electricity and
magnetism. The article is aimed at providing a new, general METHOD not found in any of
the standard texts on classical electrodynamics, which —we modestly express a hope—
should be instructive and also of didactical value.

 {\bf What are the Deliverables?} 
 
\textendash\textendash This derivation is envisioned to be instructive for standard advanced graduate coursework in relativistic electrodynamics. The major learning outcomes of this general approach are
(i) mapping of EM field tensor for a general Lorentz boost in a coordinate-free manner and (ii) to
exploit symmetry arguments to simplify tedious calculations in relativity

\section{Introduction} 
Covariance of Maxwell’s classical electrodynamics under general Lorentz transformations is
a triumph on both the counts— of consistency among all the Maxwell electrodynamics equations and of special relativity. The general (classical) Lorentz transformations of space-time coordinates encompass Lorentz-boosts, (instantaneous) proper rotations, as well as
improper transformations such as space and/or time reversals and combinations thereof. In
particular, the exclusively velocity-induced Lorentz ‘boost’ transformation connects the
space-time coordinates of an event observed in two different, relatively inertial reference frames moving with a uniform relative velocity with their Cartesian coordinate-triads
oriented parallel. Classic text-books typically present EM field transformations for idealized
setups where the underlying algebra gets drastically simplified; and then generalize the
results for arbitrary cases. Although the end result is the celebrated EM field transformation
as presented in Jackson \cite{Jackson:100964} and Landau and Lifshitz\cite{Landau:1975pou}, the most general case that involves geometric asymmetries,
arbitrary time-dependent fields, and oblique uniform relative velocities is still algebraically
tedious. In particular, it involves a laborious multiplication of a series of three 4×4 matrices
in conjunction, requiring extensive book-keeping. In this work, it will be demonstrated
that it is possible, {\it {ab initio}}, to develop an elegant general approach to EM field
transformations for the most general case employing symmetry arguments, with the end
result emerging coordinate-free. This derivation is envisioned to be instructive for
standard advanced graduate course-work in relativistic electrodynamics. The major learning
outcomes of this general approach are (i) mapping of EM field tensor for a general Lorentz
boost in a coordinate-free manner and (ii) to exploit symmetry arguments to simplify tedious
calculations in relativity. We would also like to point out that while the outcome is
admittedly way too well-known, the approach is novel, general, and to the best of our
knowledge it has not been presented hitherto in any standard text-book.

This paper is organized as follows. We first set the notation and then briefly outline the conventional textbook
approach to obtain the EM field transformation.  A naive, straightforward approach to the general EM-field transformation requires
tedious matrix multiplications and enormous book keeping.  We shall present an attractive alternative in terms of a novel approach to the general EM field transformation under arbitrary boosts, exploiting symmetry arguments.  
Here, the conventional Minkowskian set of space-time coordinates is denoted by $x^{\mu}\equiv \{x^{\mu}\}_{0}^{3}\equiv\{x^{0}=ct, x^{1}\equiv{x}, x^{2}\equiv{y}, x^{3}\equiv{z}\}$ with a Cartesian spatial part; along with
analogous connections in the primed frame. Explicitly, with $\vec{\beta}=\vec{v}/c$ and $\gamma=\sqrt{1-(\vec{\beta})^{2}}$.
The electric field $\vec{E}(x^{\mu})$ and the magnetic field $\vec{B}(x^{\mu})$ associated with given charge- and current- distributions in a frame $\Sigma$ is connected to that of primed frame $\Sigma^{'}$ as 

\begin{align}
\vec{E}^{'}=\gamma(\vec{E}+\vec{\beta}\times\vec{B})-\left(\frac{\gamma^{2}}{\gamma+1}\right)(\vec{\beta}\cdot \vec{E})\vec{\beta}\\
\vec{B}^{'}=\gamma(\vec{B}-\vec{\beta}\times\vec{E})-\left(\frac{\gamma^{2}}{\gamma+1}\right)(\vec{\beta}\cdot \vec{B})\vec{\beta}
\end{align}
where we have employed the notation as in Jackson’s classic treatise and also have chosen the Gaussian system of units so that $\vec{E}$ and $\vec{B}$ are treated on the same footing with no additional
factors of the universally constant speed of light in vacuum $c$  creeping in. Throughout, the origins of the frames are assumed to coincide at the time coordinate zero in the respective frames. 

All the accomplished texts on electricity and magnetism as well as on special theory of relativity succinctly obtain the electromagnetic field transformation equations by orienting the coordinate axes first parallel—, and then perpendicular—, to the relative uniform velocity.
For example, in his excellent and instructive text, Griffiths \cite{Griffiths:1492149} elucidates how an
electro-static field in a chosen frame $\Sigma$, manifests in another inertial Lorentz-boosted frame $\Sigma^{'}$
moving with a uniform relative velocity $\vec{v}$ with respect to $\Sigma$ , along the $x$ (positive $x$ ; also positive $x^{'}$) direction. The static charge configuration rests in $\Sigma$ , and its effects perceived in the moving frame $\Sigma^{'}$ are sought.
A uniform electrostatic field $E^{y}$ is furnished in $\Sigma$ the interior of an (in principle-) infinitely large, ideal parallel plate capacitor whose plates in $\Sigma$  bear uniform surface charge densities $-\sigma$ and $+\sigma$. The Lorentz-Fitzgerald length contraction only in the $x$ direction causes an
area-reduction as perceived in the frame $\Sigma^{'}$, whereby the surface charge densities in $\Sigma^{'}$ enhance
in accord with $\pm{\sigma^{'}}=\gamma(\pm{\sigma})$. This enhancement is directly imparted to the electric field that partially contributes (since, additionally, the magnetic effects emerge) to $E^{'y}$ to the tune of $E^{'y}=4\pi\sigma^{'}=\gamma(4\pi\sigma)=\gamma E^{y}$ as shown in Fig. 1.
Meanwhile, the electric field component along the $x$ -direction, parallel to the relative velocity, remains unaltered, as may be straightforwardly deduced if the capacitor plates were to be oriented normal to the relative boost-velocity with no areal reduction, yielding $E^{'x}=E^{x}$ , with no net magnetic contribution.
In the frame $\Sigma^{'}$, in addition to the electrostatic field, a magneto-static field also manifests, as uniform surface-currents $\vec{K}_{\pm}=\mp{\sigma{v} \hat{i}}$ also materialize since $\Sigma$ moves with $-\vec{v}$ with respect to 
$\Sigma^{'}$.

\begin{figure}[H]
    \centering
    \includegraphics[width=14cm]{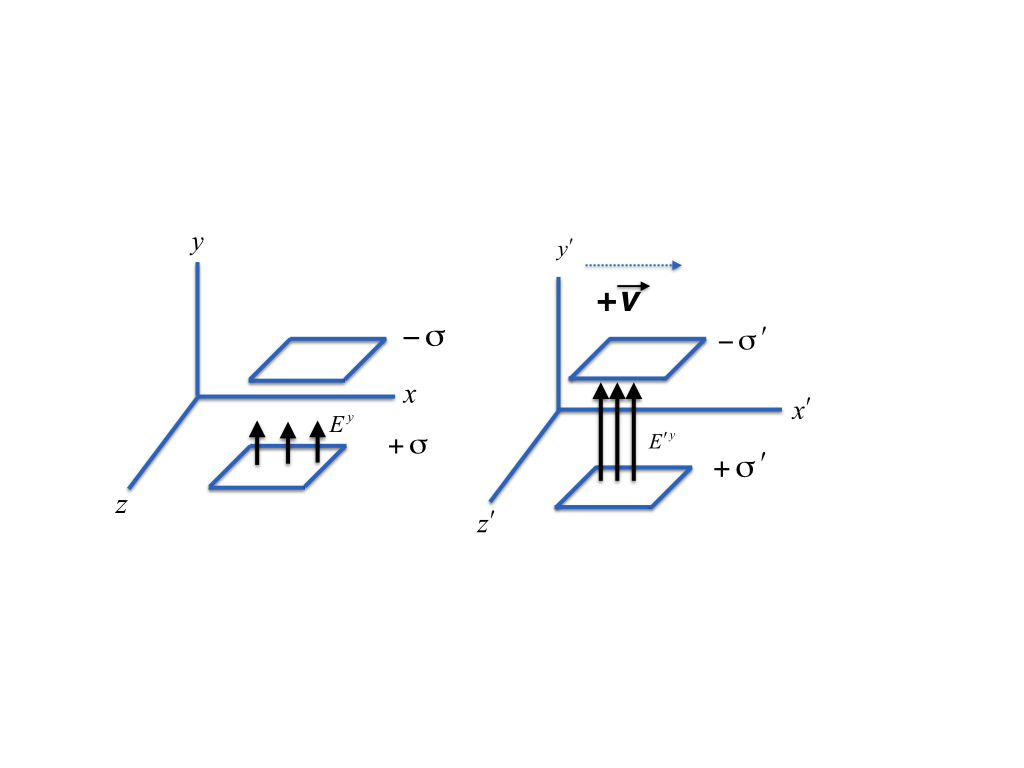}
    \caption{A schematic illustration of the transformation of the $y$-electric field (arguments
analogous to those in Griffiths \cite{Griffiths:1492149}) within the ideal parallel-plate capacitor at rest in the
frame $\Sigma$ . The Lorentz-Fitzgerald length contraction in the $x$-direction causes a reduction in area,
leading to an enhanced surface charge density, from the standpoint of $\Sigma^{'}$, relative to that in the
frame $\Sigma$ .}
   \label{fig:demo}
\end{figure}
Although we shall not elaborate here Griffiths’\cite{Griffiths:1492149} arguments as discussed clear-cut in this classic text, a third inertial frame is employed moving with a different uniform velocity in the same direction, and the velocity addition
formula in the $x$-direction is applied, obtaining the incremental component of the magneto-static field that points in the negative $z$-direction, perpendicular to the relative velocity. 

That the parallel (“ $x$ ”) component of the magnetic field is left unchanged is demonstrated
with an infinitely long cylindrical solenoid oriented in the $x$-direction with filaments wound uniformly with $n$ turns per unit axial length carrying a steady current of $I$, the internal magneto-static field engendered in $\Sigma$ in CGS units is $B^{x}=4\ \pi nI/c$. While the length contraction enhances the ‘per-unit-axial-length’ windings in $\Sigma^{'}$
are $n^{'}=\gamma n$, the current, i.e. charge flowing per unit time diminishes due to time dilation to $I^{'} = I / \gamma$ , hence the magnetic field remains still the same. Further, considering a capacitor arrangement \cite{Griffiths:1492149,CapriPanat2002} (not depicted here ) in $\Sigma$, parallel to the $x-y$ plane, and invoking once again a third Lorentz-boosted inertial frame the transformations are decisively demonstrated in a moving, relatively inertial frame, for both the electric and magnetic fields. 

With the procedure carried out to its logical end, one obtains the electromagnetic field
connections for the Cartesian field components parallel and perpendicular to the chosen relative
boost velocity in the $x$-direction:
\begin{eqnarray}
    E^{'x}&=&E^{x}\\
    E^{'y}&=&\gamma(E^{y}-\beta B^{z})\\
    E^{'z}&=&\gamma(E^{z}+\beta B^{y})\\
    B^{'x}&=&B^{x}\\
    B^{'y}&=&\gamma(E^{y}+\beta E^{z})\\
    B^{'z}&=&\gamma(E^{z}-\beta E^{y})
\end{eqnarray}
The component-fields are written with their pertinent superscripts. It may be 
noted that the Eqs. (3) and (4) make evident the intrinsic duality of the electromagnetic field
under general Lorentz transformations: $\vec{E}\rightarrow\vec{B}$ and $\vec{B}\rightarrow -\vec{E}$ hence one set directly yields the other. Eqs. (3) and (4) are readily
verifiable to be particular cases of the general formulas, Eqs. (1) and (2), with $\beta=\vec{v}/c=(v/c) \hat{i}$ in the $+x$ direction.

Akin to Griffiths’ \cite{Griffiths:1492149} deductions, Jackson \cite{Jackson:100964} in his authoritative text adroitly ‘generalizes’
the particular $x$ - Lorentz boost results; as also do Panofsky and Phillips \cite{Panofsky1962}. Another elegant
treatment by Lorrain and Corson \cite{nla.cat-vn1982088} appeals to the Lorentz force perceived in the two $x$-boosted frames that yields connections among the Cartesian components of the fields in those
frames, akin to the derivation by Resnick \cite{Resnick1968-RESITS} in his lucid introductory text on special relativity.
Remarkably, Reitz, Milford and Christy \cite{10.5555/1387035} employ the EM-field tensor and the Lorentz boosts
but only with the relative boost velocity along the chosen $x$ - direction in particular. 
Likewise, another dignified text by Landau and Lifshitz \cite{Landau:1975pou} exploits the fact that the electromagnetic $4$-Vector Potential $A^{\mu}$ under Lorentz transformations transforms exactly as the space-time coordinates  $x^{\mu}=(ct, \vec{x})$ ; consequently, $(A^{'})^{\mu}$
 in the primed frame is obtained, whose appropriate derivatives with respect to the
primed space-time coordinates yield the transformed Electric and Magnetic fields.
However, this is demonstrated only for the $x$-boosts (Equations (24.1) through (24.4) therein, pp. 62) , as it once again
becomes cumbersome to consider the oblique general Lorentz-boost.
Further, it is tacitly surmised, albeit on reasonable grounds \cite{Griffiths:1492149} that transformations established for particular
geometries should hold for time-varying cases as well. Note that the EM-field transformations
under general Lorentz-boosts, viz. Eqs. (1) and (2), are coordinate-free and embody dynamical
(time-varying charge and current densities) situations. As done in the standard texts cited above,
the results are made plausible by suitably escalating to generality the simple EM-field
connections occurring in several diversified special cases. As outlined above, these special cases
incorporate appropriately chosen ideal geometries with the boost-velocity designated along a
given Cartesian coordinate-axis.
For example, as depicted in Fig.2, an ideal parallel-plate capacitor in a frame $\Sigma$ that has
its planes intercepting all the coordinate axes (finite intercepts) can furnish in general an oblique electrostatic field with all the Cartesian components non-vanishing.
Now, if a relatively inertial frame $\Sigma^{'}$ moves (from the standpoint of the unprimed frame) with an oblique relative velocity $\vec{v}={v^{x}}{\hat{i}}+v^{y}\hat{j}+v^{z}\hat{k}$  the general Lorentz boosts bear the space-time coordinate-connections
\begin{equation}
    \vec{x}^{'}=\vec{x}+\frac{\gamma-1}{\vec{\beta}^{2}}(\vec{\beta}\cdot \vec{x})\vec{\beta}-\gamma\vec{\beta}x^{0}
\end{equation}
connecting spatial parts and temporal parts are given by
\begin{equation}
    (x^{'})^{0}=\gamma(x^{0}-\vec{\beta}\cdot \vec{x})
\end{equation}
The magnetic field from the standpoint of $\Sigma^{'}$ could also have components along all the three
Cartesian directions: thus an ideal cylindrical solenoid as noted above (not depicted herein),
oriented in a general direction in $\Sigma$ will produce all the three magnetostatic field components,
mapping onto electric and magnetic fields in the $\Sigma^{'}$ frame. Delving into ‘parallel-perpendicular’
component decompositions to extract the EM-field transformation law in such situations will
prove formidable. {\it In the most general situation, both charge and current densities will be time-dependent, a truly electro-dynamic case, further making parallel-perpendicular decompositions
aggravatingly unfeasible}.
 A legitimate question may therefore be posed: is it possible to directly obtain the EM-field
transformations under a general Lorentz-boost with oblique relative velocity between two
inertial frames, free from any encumbrances like ‘generalizing from particular instances’ and
further, to manifestly encompass dynamical situations? Moreover, how do the fields transform
under pure 3-D rotations?
\begin{figure}[H]
    \centering
    \includegraphics[width=11cm]{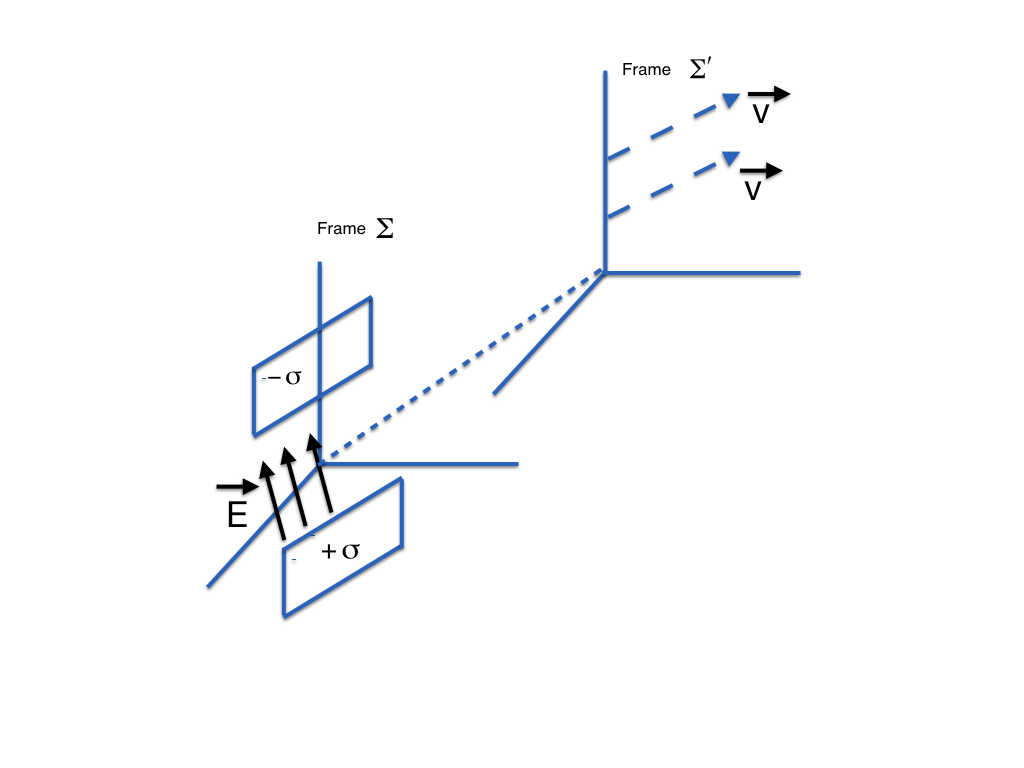}
    \caption {Two inertial frames moving with a relatively uniform oblique velocity, with an electrostatic
 field furnished with a general spatial orientation}
   \label{fig:demo2}
\end{figure}

It will now be demonstrated that these questions can indeed be answered in the affirmative,
by means of an immaculate general method rendering the oblique-boost problem
completely tractable. The key result that engenders the general electromagnetic field
transformation-equations Eq. (1) and Eq. (2) stems from how the (real) antisymmetric, rank-2
Electromagnetic Field Tensor $F\equiv F^{\mu\nu}$ defined in terms of the derivatives of the matrix elements of the four-vector potential $A\equiv \{ A^{\alpha}\}_{0}^{3}$, {\it vide} $F^{\mu\nu}=\partial^{\mu}A^{\nu}-\partial^{\nu}A^{\mu}$ maps onto its
Lorentz-transformed counterpart $F^{'}$ in another Lorentz frame. This imparts the spacetime dependence to the field tensors, in turn to the electric and magnetic field components in both the Lorentz-frames. 

In the unprimed frame, the field tensor emerges in terms of its Cartesian components thus:
\begin{equation}
\left[F^{\mu\nu}\right]=
\begin{bmatrix}
0 & -E^{1} & -E^{2} & -E^{3}\\
E^{1} & 0 & -B^{3} & B^{2}\\
E^{2} & B^{3} & 0 & -B^{1}\\
E^{3} & -B^{2} & B^{1} & 0
\end{bmatrix}
\end{equation}
Explicitly, the contravariant components of $F$ transform by the tensorial connection

\begin{equation}
    F^{'\mu\nu}=\frac{\partial{x}^{'\mu}}{\partial{x^{\alpha}}} \frac{\partial{x}^{'\nu}}{\partial{x^{\beta}}}F^{\alpha\beta}
\end{equation}
Consider the pure Lorentz-boost case first. In terms of the general Lorentz transformation matrix
$\Lambda$ transforming the space-time coordinates between the primed and the unprimed frames by 
$x^{'\mu}=\Lambda_{\nu}^{\mu}x^{\nu}$, 

\begin{align}
    F^{'\mu\nu}=&{\Lambda_{\alpha}}^{\mu}F^{\alpha\beta}{\Lambda_{\beta}}^{\nu}\\
                =& {\Lambda^{\mu}}_{\alpha}F^{\alpha\beta}\tilde{{{\Lambda}}}{^{\nu}}\,_{\beta}\nonumber
\end{align}
or, in a compact matrix notation, $F^{'}=\Lambda F \tilde{\Lambda}$; with $\tilde{\Lambda}$ denoting the transpose of $\Lambda$. Note that
$\mu$ in the above equation is the row index; and $\nu$ the column index; the summation convention being adopted throughout: be it the Greek indices such as $\mu, \nu, \alpha, \beta... $ etc. $\in \{0,1,2,3\}$ or the Roman indices 
$h, i, j, k... $ etc. $\in \{,1,2,3\}$. Moreover, the index that is the closest represents row, and the farther one, column. Meanwhile, the general Lorentz-boost with an oblique relative velocity is represented by a (real) symmetric matrix
\begin{equation}
\Lambda_{Boost}(\vec{\beta})=
\begin{bmatrix}
\gamma & -\gamma\beta^{1} & -\gamma\beta^{2} & -\gamma\beta^{3}\\
-\gamma\beta^{1} & 1+\frac{{(\gamma-1)}\beta^{1}\beta^{1}}{\vec{\beta}^{2}} & \frac{{(\gamma-1)}\beta^{1}\beta^{2}}{\vec{\beta}^{2}} & \frac{{(\gamma-1)}\beta^{1}\beta^{3}}{\vec{\beta}^{2}} \\
-\gamma\beta^{2} &\frac{{(\gamma-1)}\beta^{1}\beta^{2}}{\vec{\beta}^{2}}  & 1+\frac{{(\gamma-1)}\beta^{2}\beta^{2}}{\vec{\beta}^{2}} &\frac{{(\gamma-1)}\beta^{2}\beta^{3}}{\vec{\beta}^{2}} \\
-\gamma\beta^{3} & \frac{{(\gamma-1)}\beta^{1}\beta^{3}}{\vec{\beta}^{2}} &\frac{{(\gamma-1)}\beta^{2}\beta^{3}}{\vec{\beta}^{2}} & 1+\frac{{(\gamma-1)}\beta^{3}\beta^{3}}{\vec{\beta}^{2}}
\end{bmatrix}
\end{equation}
The subscript {“\it Boost”} will be understood hereinafter.

It is possible, in principle, to obtain the transformed fields manually by ‘brute-force’,
multiplying the three matrices as in Eq. (13) above, but this entails a complicated, daunting task to keep track and sum-up the terms, even after the components of all the 3-vectors
($\vec{E}$, $\vec{B}$, $\vec{\beta}$) are expressed as dot-products with the 
respective Cartesian unit vectors \cite{RGBrown2007}.  However, that an elegant derivation vindicating Eqs. (1) and
(2) for an arbitrary, general, oblique boost is indeed possible, is demonstrated in the next section.
It is the spirit of this article to provide a general, immaculate derivation of the electromagnetic field-transformations, as carried out below. 

\section{Derivation of EM field transformation under a general oblique Lorentz-boost }
The general derivation begins with the observation that $F^{\mu\nu}$ is a real, antisymmetric matrix while 
$\Lambda_{\beta}^{\alpha}$ (boost) is real, symmetric. Explicitly, the matrix elements of $\Lambda$ can be exhaustively represented by their values $\Lambda_{0}^{0}=\gamma$, $\Lambda_{i}^{0}=-\gamma\beta^{i}$, and $\Lambda_{j}^{i}=\delta_{j}^{i}+\frac{{(\gamma-1)}\beta^{i}\beta^{j}}{\vec{\beta}^{2}}$ and also $\Lambda_{j}^{i}=\tilde{\Lambda}^{i}\,_{j}$; which follows by  symmetry.  
$F^{00}=F^{11}=F^{22}=F^{33}=0$; $F^{0i}=-E^{i}$ $\forall{i}$; $F^{ij}=-B^{k}\varepsilon_{k}^{ij}\equiv -B^{k}\varepsilon(ijk)$ for notational convenience; and of course $F^{\mu\nu}=-F^{\nu\mu}$.
We employ here, exclusively for convenience, a ‘non-committal’ (i.e. with no ‘contra’- or ‘co’ –
variant distinction) notation for the completely antisymmetric Lévi-Civita symbol “$\varepsilon$” , since
for this symbol, particularly for three spatial dimensions, all even permutations are cyclic and the odd ones, counter-cyclic; a fact which deems that it is only the order of the symbols that matters. The context makes it evident as to the actual co- or contra- or mixed nature of the symbols. 

Next, consider a “contraction” $Q(k)$ of a symmetric symbol $S(ij)$ with the antisymmetric Lévi-Civita $\varepsilon$ , again in non-committal notations, $Q(k)=S(ij)\varepsilon(ijk)=S(ji)\varepsilon(jik)$ (dummy
repeated summation indices $i$ and $j$ that appear take on the values = 1, 2, and 3). However, $S(ij)=S(ji)$ $\forall i,j$; hence it follows that $Q(k)=\frac{1}{2}S(ij)\left[\varepsilon(ijk)+\varepsilon(jik)\right]=0$, {\it vide} the
antisymmetry of $\varepsilon$; meaning that a generic contraction of antisymmetry and symmetry sums to zero.

With these preliminaries, consider now the matrix elements of $F^{'}$: $(F)^{'\alpha\beta}=\left(\Lambda F \tilde{\Lambda}\right)^{\alpha\beta}=\left(\Lambda F {\Lambda}\right)^{\alpha\beta}$. Setting $\alpha=0$ and $\beta\equiv i$ in conjunction with symmetry properties of $\Lambda$, we can show that 
\begin{align}
    (F)^{'0i}&= -E^{'i}\nonumber\\ 
             &=\Lambda_{\alpha}^{0}F^{\alpha\beta}\Lambda_{\beta}^{i}\nonumber \\ 
             &=\Lambda_{0}^{0}F^{0k}\Lambda_{k}^{i}+\Lambda_{g}^{0}F^{gh}\Lambda_{h}^{i}+\Lambda_{k}^{0}F^{k0}\Lambda_{0}^{i}+\Lambda_{0}^{0}F^{00}\Lambda_{0}^{i}
\end{align}

The last term vanishes ($\because{F^{00}=0}$), and after inserting the respective matrix elements, one is led to 

\begin{align}
    (-E')^{i}&=\gamma(-E)^{k}\left[\delta_{i}^{k}+\frac{{(\gamma-1)}\beta^{i}\beta^{k}}{\vec{\beta}^{2}}\right]-(\gamma\beta^{g})\varepsilon(fgh)(-B^{f})\left[\delta_{h}^{i}+\frac{{(\gamma-1)}\beta^{h}\beta^{i}}{\vec{\beta}^{2}}\right]\\
    &+(-\gamma\beta^{k})(+E^{k})(-\gamma\beta^{i})+0\nonumber
\end{align}
Recognizing that $E^{k}\beta^{k}=\vec{E}\cdot\vec{\beta}$; and collecting and regrouping the terms in the above equation, the equation simplifies to

\begin{align}
    (-E')^{i}&=\gamma E^{i}+\left[\frac{\gamma(\gamma-1)}{{\vec{\beta}^{2}}}-\gamma^{2}\right]\left(\vec{E}\cdot\vec{\beta}\right)\beta^{i}+\gamma\beta^{g}B^{f}\varepsilon(fgi)+\frac{(\gamma-1)\beta^{i}\beta^{g}B^{f}\beta^{h}\varepsilon(fgh)}{\vec{\beta^{2}}}
\end{align}
The last term on the right completely vanishes as it embodies a ‘contraction’ of a product $(\beta^{g}\beta^{f})$ that is completely symmetric under interchange of indices, with the Lévi-Civita symbol $\varepsilon(fgh)$ that is anti-symmetric.
The third term on the right contains a factor $\beta^{g}B^{f}\varepsilon(fgi)$ which is exactly the component $-\gamma(\vec{\beta}\times\vec{B})^{i}$. Therefore, the above equation with $\beta^{2}=1-1/\gamma^{2}$ gives desired complete 3-vector form

\begin{equation}
    \vec{E}^{'}=\gamma(\vec{E}+\vec{\beta}\times\vec{B})-\left(\frac{\gamma^{2}}{\gamma+1}\right)(\vec{\beta}\cdot \vec{E})\vec{\beta}
\end{equation}
which precisely is Eq. (1) for the transformation of the electric field $\vec{E}(\vec{x},t)$ in the frame $\Sigma$ into a
general Lorentz-boosted frame $\Sigma^{'}$ moving with respect to the former, with an arbitrary, oblique $\vec{v}=\vec{\beta}c$.

As for the magnetic field, one may simply invoke the dual electromagnetic field tensor
$F_{dual}$ which is defined as $F_{dual}^{\mu\nu}=\varepsilon^{\mu\nu\alpha\beta}F_{\alpha\beta}/2$, readily obtained from $F^{\mu\nu}$ with the replacements $\vec{E}\rightarrow\vec{B}$ and $\vec{B}\rightarrow \vec{-E}$ carried out. The dual tensor is also antisymmetric and satisfies an analogous connection ${F^{'}}_{dual}=\Lambda F_{dual} \tilde{\Lambda}$.
This amounts to directly replacing $\vec{E}\rightarrow\vec{B}$ and $\vec{B}\rightarrow \vec{-E}$ in Eq. (1) leading to Eq. (2)
\begin{equation}
    \vec{B}^{'}=\gamma(\vec{B}-\vec{\beta}\times\vec{E})-\left(\frac{\gamma^{2}}{\gamma+1}\right)(\vec{\beta}\cdot \vec{B})\vec{\beta}.\nonumber\\
\end{equation}
Albeit by a slightly devious route, it is possible to obtain the magnetic field
transformation directly from $F^{\mu\nu}$ without invoking its dual tensor. This instructive
exercise is carried out in detail in the Appendix.
Further, the electric and magnetic fields can in general be space- and time-dependent, and not only static ones in suitably contemplated special inertial frames. 

Note that the Lorentz-transformation is a proper, real, but a non-orthogonal one. On the other hand,
Classical Lorentz group also incorporates pure, 3-Dimensional proper rotations. It would prove rewardingly instructive to examine how the electromagnetic fields transform under this physical transformation, as discussed below.

\section{HOW DOES THE ELECTROMAGNETIC FIELD TRANSFORM UNDER SPACE-
 ROTATIONS ? }
 For transformation of the electric and magnetic fields
 under three-dimensional (3-D) proper rotations, we follow the “axis-angle” representation of Rotation as established in the revered classic reference by Goldstein, Safko and Poole \cite{Goldstein2014} where, a 3-D spatial vector $\vec{x}$ in a frame with its triad of axes fixed in space, that is rotated to its new orientation $\vec{x^{'}}$ can be
regarded as being caused by a  pure spatial {\it active} rotation around a given, fixed axis $\hat{n}$ by a given angle $\Phi$:
\begin{equation}
    \vec{x^{'}}=\hat{R}_{sp}(\hat{n},\Phi)\vec{x}
\end{equation}
Or in terms of matrix elements 
\begin{equation}
    x^{'i}={R}_{sp,j}^{i}x^{j}   \quad\forall{i}=1,2,3
\end{equation}
Explicitly, 
\begin{equation}
    \vec{x^{'}}=\hat{n}(\hat{n}\cdot\vec{x})(1-\cos(\Phi))+\vec{x}\cos(\Phi)
    -(\vec{x}\times \hat{n})\sin(\Phi)
\end{equation}
If, instead, the vector is rigidly {\it clamped} at the origin, and the {\it axes} are rotated, constituting a {\it passive} rotation, the transformation law simply has sign of the last term reversed: 
\begin{equation}
    \vec{x^{'}}=\hat{n}(\hat{n}\cdot\vec{x})(1-\cos(\Phi))+\vec{x}\cos(\Phi)
    +(\vec{x}\times \hat{n})\sin(\Phi)
\end{equation}
as in Goldstein {\it et al.} \cite{Goldstein2014}, and also elaborated elsewhere \cite{LectureUCSC2016}.

Intuitively, the electric and magnetic fields considered as 3-D vectors, should follow the 3-vector $\vec{x}$. However, there is a question: Do the vectors $\vec{E}$ and $\vec{B}$ that are conventionally not $3$-vector parts of some Lorentz-$4$-vectors, follow active or passive rotations? More
importantly, why so? For an active rotation of the components of the vector, we follow the connection (28), as expressed above. In general, any {\it instantaneous Proper 3-D Spatial Rotation} is an element of the group $SO(3)$, where, by its veritable construction, the time coordinate remains unchanged, hence $x^{'0}\equiv ct^{'}=ct\equiv x^{0}$. In particular, for an active instantaneous 3-D spatial proper rotation, the matrix elements of the $4\times4$ matrix that belongs to a
Lorentz transformation in a general sense, viz. $R\equiv R(\hat{n},\Phi)$ has an explicit form
\begin{equation}
R(\hat{n},\Phi)=
   \begin{bmatrix}
    1 & {\mathcal{O}}_{1\times3}\\
    {\mathcal{O}}_{3\times1} & R_{sp}(\hat{n},\Phi)
    \end{bmatrix}
\end{equation}
where ${\mathcal{O}}_{1\times3}=[0,0,0]$ and ${\mathcal{O}}_{3\times1}$ is a column of zeros.
Here the superscripts denote the respective Cartesian components of the fixed unit vector $\hat{n}$, that defines the axis of rotation. Note that $R_{\mu}^{0}=R_{0}^{\mu}=\delta^{0}_{\mu}=(\delta^{\mu}_{0})$, which also embodies $R^{0}_{i}=R^{i}_{0}=0$ $\forall {i=1,2,3}$; since in the zeroth row and also in the zeroth column, only $R^{0}_{0}=1$ while all the other elements vanish. The transformation matrix is thus manifestly “block-diagonal” into the $1\times1$ “time”-sector containing the solitary non-vanishing element $R^{0}_{0}=1$ and
the $3\times3$ “spatial” sector, executing any instantaneous spatial transformations. We label the spatial-sector in the transformation
Eq. (30) by $R_{sp}\equiv [R^{i}_{sp,j}]$ with $R_{sp}\in{SO(3)}$
 effecting active 3-D rotations through the orthogonal matrix of transformation: 
 
 \begin{align}
R_{sp}(\hat{n},\Phi)=
    \begin{bmatrix}
      n^{1}n^{1}(1-\cos(\Phi))+\cos(\Phi) & n^{1}n^{2}(1-\cos(\Phi))-n^{3}\sin(\Phi) & n^{1}n^{3}(1-\cos(\Phi))+n^{2}\sin(\Phi)\\
      n^{2}n^{1}(1-\cos(\Phi))+n^{3}\sin(\Phi) & n^{2}n^{2}(1-\cos(\Phi))+\cos(\Phi) & n^{2}n^{3}(1-\cos(\Phi))-n^{1}\sin(\Phi)\\
      n^{3}n^{1}(1-\cos(\Phi))-n^{2}\sin(\Phi) & n^{3}n^{2}(1-\cos(\Phi))+n^{1}\sin(\Phi) & n^{3}n^{3}(1-\cos(\Phi))+n^{3}\cos(\Phi)
    \end{bmatrix}
\end{align}
Symbolically, $R^{i}_{sp,j}$ the matrix elements incarnate as:
\begin{equation}
    R^{i}_{sp,j}=(n^{i})(n^{j})(1-\cos(\Phi))+\delta^{i}_{j}\cos(\Phi)-\epsilon^{i}_{jk}n^{k}\sin(\Phi)
\end{equation}
Note here that on the right hand side, only the {\it values} of the quantities are to be taken into account. The superscripts on the unit vector of course mean $n^{1}=n^{x}$ the $x$-component, etc. [Although neither used nor alluded to herein, it is still maintained that $n_{i}=-n^{i}$; for consistency, in view of the signature of the flat Minkowskian metric $diag(1, -1, -1, -1)$ ]. For practical
purposes, once again, suspending the contra- or co- variant symbolism on the right-side for the Kronecker-$\delta$ and the Lévi-Civita-symbol $\varepsilon$, 
\begin{equation}
    R^{i}_{sp,j}=(n^{i})(n^{j})(1-\cos(\Phi))+\delta(ij)\cos(\Phi)-\varepsilon(ijk)n^{k}\sin(\Phi)
\end{equation}
We have the Lorentz-transformation connection, for pure rotations, compactly expressed as a matrix product in terms of the larger, Lorentz-group element matrix $R$ :
\begin{equation}
    F^{'}=RF\tilde{R}
\end{equation}
With the convention, as before, that the index closest to the matrix-symbol is the row index, whence $(\tilde{R})^{\nu}_{\beta}=R_{\beta}\,^{\nu}$, leading, for the electric field transformation

\begin{align}
    (F)^{'0i}&= -E^{'i}\\
             &={R_{\alpha}}^{0}F^{\alpha\beta}{R_{\beta}}^{i} \nonumber\\
             &={R_{0}}^{0}F^{0j}{R_{j}}^{i}+\cancelto{0}{R_{g}}^{0}F^{gh}{R_{h}}^{i}+\cancelto{0}{{R_{k}}^{0}F^{k0}{R_{0}}^{i}}+\cancelto{0}{R_{0}}^{0}F^{00}{R_{i}}^{0}\nonumber
\end{align}
Note that, since ${R_{j}}^{i}={R_{sp,j}}^{i}$, this yields 
\begin{equation}
    (F)^{'0i}=-(E^{'})^{i}=F^{0j}{R_{sp,j}}^{i}=-E^{j}{R_{sp,j}}^{i}
\end{equation}
simplifying into 
\begin{equation}
    E^{'i}=E^{j}{R_{sp,j}}^{i}
\end{equation}
This connection immediately implies that the transformed electric field under instantaneous pure 3-D spatial rotations will not have any magnetic field counterpart of the unprimed frame, and by the electromagnetic field duality, the magnetic field will also not have the unprimed electric field. This is plausible, since, for example, an electro-{\it static} situation in one frame will also continue to be a static one in a frame that has its axes rotated; similarly for a magneto-{\it static} effect there will be no relevance of any electric field. This situation is unlike that for the pure
Lorentz boosts, since only when there is relative uniform “boost” motion between two inertial frames the dynamics would materialize that imparts, for the electric field transformation, over and above the obvious electric effect, an incremental magnetic effect from the standpoint of the moving frame, and reciprocally for the magnetic field. Note the subtle difference that the transformation matrix appearing on the extreme right is actually the {\it transpose} of $R_{sp}$ , also an orthogonal matrix, exactly as appearing in a passive (vector-fixed, axes-rotating) transformation. 
Thus,
\begin{equation}
    \vec{E^{'}}=\hat{n}(\hat{n}\cdot\vec{E})(1-\cos(\Phi))+\vec{E}\cos(\Phi)
    +(\vec{E}\times \hat{n})\sin(\Phi)
\end{equation}
Exactly the same argument holds for the magnetic field when one applies the duality 
$\vec{E}\rightarrow\vec{B}$ and $\vec{B}\rightarrow-\vec{E}$

\begin{equation}
    \vec{B^{'}}=\hat{n}(\hat{n}\cdot\vec{B})(1-\cos(\Phi))+\vec{B}\cos(\Phi)
    +(\vec{B}\times \hat{n})\sin(\Phi)
\end{equation}
To confirm that the transformation Eqs. (38) and (39) are legitimate, a sufficient condition that is easily verifiable is that the quantity $\vec{E}\cdot\vec{B}$ be truly a Lorentz-invariant:
\begin{align}
    \vec{E^{'}}\cdot\vec{B^{'}}&=(\vec{E}\cdot\hat{n})(\vec{B}\cdot\hat{n})\left[(1-\cos(\Phi))^{2}+2\cos(\Phi)(1-\cos(\Phi))\right]+\vec{E}\cdot\vec{B}\cos^{2}(\Phi)\\ &+(\vec{E}\times\hat{n})\cdot(\vec{B}\times\vec{n})\sin^{2}(\Phi)
     +\left[\vec{E}\cdot(\vec{B}\times\hat{n})+\vec{B}\cdot(\vec{E}\times\hat{n})\right]\sin(\Phi)\cos(\Phi) \nonumber
\end{align}
Noting that $(\vec{E}\times\hat{n})\cdot(\vec{B}\times\vec{n})=\vec{E}\cdot\vec{B}-(\vec{E}\cdot\hat{n})(\vec{B}\cdot\hat{n})$, in conjunction with $\vec{E}\cdot(\vec{B}\times\hat{n})=-\vec{E}\cdot(\vec{B}\times\hat{n})=-\vec{B}\cdot(\vec{E}\times\hat{n})$ so that the $\vec{E}\cdot(\vec{B}\times\hat{n})+\vec{B}\cdot(\vec{E}\times\hat{n})=\vec{0}$
 ; whence the right side of Eq. (40) simplifies itself to yield 
 \begin{equation}
     \vec{E^{'}}\cdot\vec{B^{'}}=\vec{E}\cdot\vec{B}
 \end{equation}
 genuinely a Lorenz-invariant, as required. Further, under pure spatial proper rotations, the magnitude of the electric field turns out to be is invariant: 
 \begin{align}
     (\vec{E}^{'})^{2}&=(\vec{E}\cdot\hat{n})^{2}\left[(1-\cos(\Phi)^{2}+2\cos(\Phi)(1-\cos(\Phi))\right] 
     +(\vec{E})^{2}\cos^{2}(\Phi) \\
     & +(\vec{E}\times\hat{n})\cdot(\vec{E}\times\hat{n})\sin^{2}(\Phi)\nonumber \\
     & =(\vec{E}\cdot\hat{n})^{2}(1-\cos^{2}(\Phi))+\left[\vec{E}^{2}-(\vec{E}\cdot\hat{n})^{2}\right]\sin^{2}(\Phi) \nonumber\\
     & = (\vec{E})^{2}
 \end{align}
Likewise, $\vec{B}^{2}=\vec{B'}^{2}$, hence, the magnitudes of the electric and magnetic fields are individually unchanged under pure space-rotations; evidently, the difference
$(\vec{E^{'}}^{2}-\vec{B^{'}}^{2})=(\vec{E}^{2}-\vec{B}^{2})$ also
becomes Lorentz-invariant, per the stringent requirement of the general Lorentz-transformations. These attributes validate the electromagnetic field transformations Eqs. (38) and (39) for instantaneous proper 3-D rotations.

\section{Concluding remarks}
This article presents a direct derivation of how a general electromagnetic field transforms
under arbitrary Lorentz-boosts. The present derivation assumes neither any preferred axes, nor any preferred symmetries, nor any deliberately furnished, ideal, static (in a given Lorentz frame) electric or magnetic fields with ideal apparatuses, and is not marred by requiring any ‘suitable’ generalizations, unlike as done in
standard texts on electrodynamics. The treatment has a direct bearing exclusively on the subtle
mapping of the electromagnetic field tensor induced by a general Lorentz boost transformation
with any given arbitrarily oriented relative boost-velocity. The end result manifests in a natural,
coordinate-free form where the electric and magnetic fields appear together as a composite
entity, that is, the electromagnetic field. As to the pure, instantaneous, proper 3-D Spatial
Rotations, the electric and magnetic fields follow the passive rotation transformations, and
remarkably, do not ‘mix’ up: electric field in one frame maps onto exclusively an electric field in
the rotated frame, the magnetic field following similar suit. It is earnestly felt that the present
work should buttress the standard textbook derivations through cogent, rigorous physical arguments. 

\section{Acknowledgments}
RKP is indebted to Mr. Todd W. Wegner for asking thought-provoking questions that
prompted the authors to undertake the present instructive study. A visiting professorship to
Tulane University, New Orleans, LA, U.S.A., where which this work was initiated a while ago;
as well as the current honorary adjunct professorship by the Physics Department, S.P. Pune
University, MH, India are both gratefully acknowledged by RKP. 

\appendix
\section{DIRECT DERIVATION OF THE MAGNETIC FIELD TRANSFORMATION WITHOUT
INVOKING THE DUAL ELECTROMAGNETIC FIELD TENSOR
}
It also would prove instructive to perform the exercise directly, without any reference to the dual field tensor, as follows: note that the ‘spatial’ part of the electromagnetic field tensor is expressed as 

\begin{align}
    (F^{'})^{ij}&= \varepsilon(ijk)(-B^{'})^{k}\nonumber\\
                &=\Lambda_{\alpha}^{i}F^{\alpha\beta}\Lambda_{\beta}^{j}\nonumber \nonumber\\
                &=\Lambda_{0}^{i}F^{00}\Lambda_{0}^{j}+\Lambda_{0}^{i}F^{0l}\Lambda_{l}^{j}+\Lambda_{l}^{i}F^{lm}\Lambda_{m}^{j}+\Lambda_{l}^{i}F^{l0}\Lambda_{0}^{j};
\end{align}
with the first term vanishing, since $F^{00}=0$. Please note that $\Lambda=\tilde{\Lambda}$ for boosts. A direct substitution for the matrix elements yields, on the right side 

\begin{align}
    \varepsilon(ijk)(-B^{'})^{k}&=0+(\gamma\beta^{i})(-E)^{l}\left[\delta_{l}^{j}+\frac{{(\gamma-1)}\beta^{j}\beta^{l}}{\vec{\beta}^{2}}\right]\nonumber\\
    &+\left[\delta_{l}^{j}+\frac{{(\gamma-1)}\beta^{j}\beta^{l}}{\vec{\beta}^{2}}\right](-B^{p})\varepsilon(lmp)\left[\delta_{m}^{j}+\frac{{(\gamma-1)}\beta^{j}\beta^{m}}{\vec{\beta}^{2}}\right]\nonumber \nonumber\\&+\left[\delta_{l}^{i}+\frac{{(\gamma-1)}\beta^{i}\beta^{l}}{\vec{\beta}^{2}}\right](E^{l})(-\gamma\beta^{j})
\end{align}
Next, we open up the brackets, giving

\begin{align}
    \varepsilon(ijk)(-B^{'})^{k}&=(\gamma\beta^{i})(E^{j})+\cancel{
    \frac{\gamma(\gamma-1)\beta^{i}\beta^{j}(E^{l}\beta^{l})}{\vec{\beta}^{2}}}\nonumber\\ 
    &-\delta_{l}^{i}\delta_{m}^{j}(B^{p})\varepsilon(lmp)-
    \delta_{l}^{i}(B^{p})\epsilon(lmp)\beta^{m}\beta^{j}\frac{(\gamma-1)}{\vec{\beta}^{2}} \nonumber\\  
    &-\delta_{m}^{j}(B^{p})\epsilon(lmp)\beta^{i}\beta^{l}\frac{(\gamma-1)}{\vec{\beta}^{2}} -\varepsilon(lmp)\beta^{i}\beta^{l}\beta^{m}\beta^{j}\left[\frac{(\gamma-1)}{\vec{\beta}^{2}}\right]^{2}(B^{p})-\gamma\beta^{j}(E^{i})\nonumber\\  
    & -\cancel{\frac{\gamma(\gamma-1)\beta^{i}\beta^{j}(E^{l}\beta^{l})}{\vec{\beta}^{2}}}
\end{align}

The expression on the right hand side simplifies itself, after recognizing that the second and the eighth terms cancel out, and the sixth term vanishes since it embodies contraction of the anti-symmetric tensor $\varepsilon(lmp)$  with a symmetric product 
$\beta^{l}\beta^{m}$, in the indices $l$ and $m$. Whence, 

\begin{align}
    (B^{'})^{k}\varepsilon(ijk)&=(B^{p})\varepsilon(ijp)-\gamma\beta^{i}(E^{j})+(B^{p})\varepsilon(imp)\beta^{m}\beta^{j}\frac{(\gamma-1)}{\vec{\beta}^{2}}\nonumber\\  
    & + (B^{p})\varepsilon(ljp)\beta^{i}\beta^{l}\frac{(\gamma-1)}{\vec{\beta}^{2}}+\gamma\beta^{j}(E^{i}) 
\end{align}

We now multiply both the sides by $\varepsilon(ijq)$ (summation carried over the indices $i$, $j$ and $p$ is tacit, as they are all repeated); and use the identities relating to contractions of the symbols, viz.

\begin{align}
    \epsilon(ijk)\epsilon(ijq)&=2\delta_{q}^{k}\\
    \epsilon(ijk)\epsilon(ipq)&=\delta_{p}^{j}\delta_{q}^{k}-\delta_{q}^{j}\delta_{p}^{k}\nonumber
\end{align}
Simplification of the third and fourth terms in Eq. (A4), that merits elucidation, occurs as follows: 
Let us consider these terms in conjunction. After multiplication with $\epsilon(ijq)$ throughout,

\begin{align*}
    (B^{p})\varepsilon(imp)\varepsilon(ijq)\beta^{m}\beta^{j}\frac{(\gamma-1)}{\vec{\beta}^{2}}+(B^{p})\varepsilon(ljp)\varepsilon(ijq)\beta^{i}\beta^{l}\frac{(\gamma-1)}{\vec{\beta}^{2}}\\
    =(B^{p})\left[\delta_{p}^{q}\delta_{m}^{j}-\delta_{m}^{q}\delta_{j}^{p}\right]\beta^{m}\beta^{j}\frac{(\gamma-1)}{\vec{\beta}^{2}}+
    (B^{p})\left[\delta_{p}^{q}\delta_{l}^{i}-\delta_{l}^{q}\delta_{i}^{p}\right]\beta^{p}\beta^{i}\frac{(\gamma-1)}{\vec{\beta}^{2}}\nonumber;
\end{align*}
where, we have exploited the symmetry of the Kronecker-$\delta$ symbol in its indices. 
Consequently, the above combination reduces to 
\begin{equation}
    (B^{q})\beta^{m}\beta^{m}\frac{(\gamma-1)}{\vec{\beta}^{2}}+
    (B^{p})\beta^{p}\beta^{q}\frac{(\gamma-1)}{\vec{\beta}^{2}}+
    (B^{q})\beta^{i}\beta^{i}\frac{(\gamma-1)}{\vec{\beta}^{2}}-
    (B^{p})\beta^{p}\beta^{q}\frac{(\gamma-1)}{\vec{\beta}^{2}}\nonumber
\end{equation}
Now $\beta^{m}\beta^{m}=\vec{\beta}^{2}$ and $\beta^{p}B^{p}=\vec{\beta}\cdot\vec{B}$, a scalar product of the 3-vectors, which renders the above expression into
\begin{align}
  2(B^{q})\beta^{m}\beta^{m}\frac{(\gamma-1)}{\vec{\beta}^{2}}-
  2(B^{p})\beta^{p}\beta^{q}\frac{(\gamma-1)}{\vec{\beta}^{2}}
  =2B^{q}(\gamma-1)-2(\vec{\beta}\cdot\vec{B})\frac{(\gamma-1)}{\vec{\beta}^{2}}\nonumber
\end{align}

Terms involving the electric field add up by the anti-symmetry of the
Lévi-Civita symbol: $-2\gamma\beta^{i}(E^{j})\epsilon(ijq)$, which readily identifies itself with $-2\gamma(\vec{\beta}\times\vec{E})^{q}$.

All the above considerations lead to 
\begin{equation}
    2(B^{'})^{q}=2(B)^{q}-2\gamma(\vec{\beta}\times\vec{E})^{q}+2B^{q}(\gamma-1)-2\frac{(\gamma-1)}{\vec{\beta}^{2}}(\vec{\beta}\cdot\vec{B})\vec{\beta}
\end{equation}
For any arbitrary component “ q ” of the magnetic field, $\vec{B}$. 
Upon canceling out the common factor of $2$, trading off ${\vec{\beta}}^{2}$ in terms of $\gamma$ ; and escalating the equation to its rightful 3-vector status one arrives at the famous result 

\begin{equation}
   \vec{B}^{'}=\gamma(\vec{B}-\vec{\beta}\times\vec{E})-\left(\frac{\gamma^{2}}{\gamma+1}\right)(\vec{\beta}\cdot \vec{B})\vec{\beta} \nonumber
\end{equation}
which is exactly Eq. (2), for the transformation of the magnetic field.

\clearpage
\bibliography{main.bib}

\end{document}